\documentclass{ws-ijmpa}

\begin{document}

\markboth{B. Desplanques}{Forms of Relativistic Kinematics}

\catchline{}{}{}{}{}

\title{NUCLEON AND PION FORM FACTORS IN DIFFERENT FORMS OF RELATIVISTIC
KINEMATICS}

\author{B. DESPLANQUES}
\address{Laboratoire de Physique Subatomique et de Cosmologie (UMR
IN2P3/CNRS-UJF-INPG), F-38026 Grenoble Cedex, France}

%%%%%%%%%%%%%%%%%%%%%%%%%%%%%%%%%%%%%%%%%%%%%%%%%%%%%%%%%%%%
% You may repeat \author \address as often as necessary    %
%%%%%%%%%%%%%%%%%%%%%%%%%%%%%%%%%%%%%%%%%%%%%%%%%%%%%%%%%%%%

\maketitle

%\begin{history}
%\received{DAY MONTH YEAR}
%\revised{DAY MONTH YEAR}
%\end{history}

\begin{abstract}
Calculations of form factors in different forms of relativistic kinematics
are presented. They involve the instant, front and point forms. 
In the two first cases, different kinematical conditions are considered 
while in the latter case, both a Dirac-inspired approach and 
a hyperplane-based one are incorporated in our study. Numerical results 
are presented for the pion form factors with emphasis on both the low 
and high $Q^2$ range. A new argument is presented, explaining why some 
approaches do considerably much better than other ones when
only a single-particle current is considered.
\end{abstract}

\keywords{03.65.Pm; 21.45.+v; 13.40.Gp}

%%%%%%%%%%%%%%%%%%%%%%%%%%%%%%%%%%%%%%%%%%%%%%%%%%%%%%%%%%%%
% The main text of your paper	begins here			     %
%%%%%%%%%%%%%%%%%%%%%%%%%%%%%%%%%%%%%%%%%%%%%%%%%%%%%%%%%%%%

%%%%%%%%%%%%%%%%%%%%%%%%%%%-1111111111111-%%%%%%%%%%%%%%%%%%%%%%%%%%%%%%%%%%%%%%%
%%%%%%%%%%%%%%%%%%%%%%%%%%%%%%%%%%%%%%%%%%%%%%%%%%%%%%%%%%%%%%%%%%%%%%%%%%%%%%%%%
%%%%%%%%%%%%%%%%%%%%%%%%%%%%%%%%%%%%%%%%%%%%%%%%%%%%%%%%%%%%%%%%%%%%%%%%%%%%%%%%%
\section{Introduction}
The knowledge of hadron form factors, especially for the nucleon and 
the pion ones, represents an important source of information about 
the structure of the systems under consideration. By varying the momentum 
transfer, large as well as small distances can be explored, allowing one 
to learn about hadronic physics in the perturbative and non-perturbative 
regimes of QCD and its modelization. Involving large momentum transfers, 
the above study supposes that a reliable implementation of relativity 
is made. This is mandatory if some information about the hadronic 
structure is to be looked for from experiments. 

There are many ways to implement relativity in the description of properties 
of a few-body system. The most ambitious one is based on field theory but, 
at present, its use for the nucleon form factor is hardly conceivable. 
A quite different approach involves relativistic quantum mechanics (RQM), 
which contrary to the previous one, assumes a fixed number of degrees 
of freedom. Less fundamental, this one is however more adapted when 
a modelization of hadrons from constituent particles is used, as most 
often done. Following Dirac,\cite{Dirac:1949cp} many approaches along 
these lines have been proposed depending on the symmetry properties 
of the hypersurface on which physics is described. This reflects in
the construction of the Poincar\'e group generators, which drop accordingly 
into dynamical and kinematical operators.

When calculating properties such as form factors, all approaches should 
converge to a unique answer but, of course, some may be more convenient 
in that the bulk contribution is produced by a one-body current. 
In other ones, large contributions from two- or many body-currents 
may be required, possibly obscuring conclusions about the physics. 
This requires that independent studies be performed to establish 
the respective advantage of various approaches by comparing their predictions. 

Those studies that will be presented here have largely been motivated 
by the successful description of the nucleon form factors in the 
``point-form'' approach\cite{Wagenbrunn:2000es} while a standard 
front-form one\cite{Cardarelli:1995dc} is failing in the same conditions. 
Adding to this puzzling
situation, it is noticed that accounting for the well known physics 
underlying the vector-meson-dominance phenomenology, ignored in the former 
case, would reduce the discrepancy in the latter one. For the present 
purpose however, we will consider a system simpler than the nucleon, 
namely the pion. Apart from the fact that there is an evident logics 
in considering systems with increasing complexity, the smallness 
of the pion mass in comparison with the sum of the constituent masses 
turns out to considerably enhance the differences between various approaches. 
This can contribute to sharpen the conclusions. 

The plan of the paper is as follows. In the second section, we precise the
ingredients entering the calculation of form factors in different kinematics 
of relativitic quantum mechanics (RQM): instant, front and point forms.
For each approach different cases, described in the text, are considered.  
Results for both the charge and scalar pion form factors are presented 
and discussed in the third section. Some attention is given to their 
asymptotic behavior. The fourth section is devoted to the conclusion 
where the role of the  space-time translation invariance is evoked. 
Due to limited space, we skip many details and refer 
to published works for 
them.\cite{Amghar:2002jx,Desplanques:2004rd,Desplanques:2004sp}
%%%%%%%%%%%%%%%%%%%%%%%%-22222222222222-%%%%%%%%%%%%%%%%%%%%%%%%%%%%%%%%%%%%%%%
%%%%%%%%%%%%%%%%%%%%%%%%%%%%%%%%%%%%%%%%%%%%%%%%%%%%%%%%%%%%%%%%%%%%%%%%%%%%%%%
%%%%%%%%%%%%%%%%%%%%%%%%%%%%%%%%%%%%%%%%%%%%%%%%%%%%%%%%%%%%%%%%%%%%%%%%%%%%%%%
%%%%%%%%%%%%%%%%%%%%%%%%%%%%%%%
\section{Different forms of relativistic quantum mechanics: a few points} 
\label{sec2}
\begin{figure}[htb] 
\hspace*{3cm}\epsfig{file=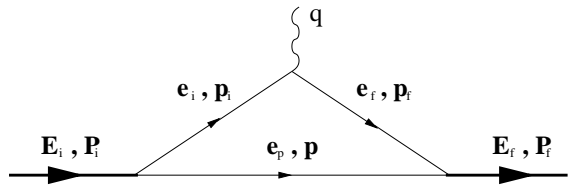, width= 7.5cm} 
\caption{Contribution to the form factor in the single-particle approximation}
\label{fig1}
\end{figure}
In order to calculate form factors of a given system, two ingredients 
are needed: the relation between the momenta of its constituents and 
the total momentum, which  characterizes each  RQM approach (see kinematics 
in Fig. \ref{fig1}), and a solution of a mass operator, 
which can be chosen as independent of the approach. 
They are successively discussed in the following.

For the two-body system of interest here, the relation between the momenta 
of its constituents and the total momentum takes a unique form. This one reads: 
\begin{equation}
\vec{p}_1+\vec{p}_2-\vec{P} = \frac{\vec{\xi}}{\xi^0}\;(e_1+e_2-E_P),
\label{eq1}
\end{equation}
where the 4-vector, $\xi^{\mu}$, is representative of the symmetry 
properties (if any) evidenced by the hypersurface which physics is described on. 
Accordingly, the Poincar\'e group
generators, $P^{\mu} \;(P^0,\,\vec{P})$ and $M^{\mu \nu}\;(\vec{K},\,\vec{J})$ 
drop into dynamical or kinematical ones.\cite{Keister:1991sb}
This character  together with the 4-vector $\xi^{\mu}$ are precised below:\\
- instant form: 
$t =\tau,\;\;\;\xi^0=1, \;\;\;  \;\vec{\xi}=0\,$;\\
\hspace*{1cm}dynamical: $P^0,\;\vec{K}, \;\;\;$ kinematical:
$\vec{P},\;\vec{J}\,$,\\
- front form: 
$t-\vec{n}\cdot\vec{x}=\tau,\;\;\;\xi^0=1, \;\;\;  \;\vec{\xi}=\vec{n}\,,
$ \\
\hspace*{6mm}where $\vec{n}$ is a unit vector with a fixed direction, generally 
chosen opposite to the \hspace*{6mm}z-axis orientation ($\xi^2=0$);\\
\hspace*{1cm}dynamical: $P^0\!-\!P^z,\;J_{\perp},\;\;\;$ kinematical: 
$P^0\!+\!P^z,\;P_{\perp},\;J^z,\;K^z,
\; K_{\perp} \!-\! \hat{z} \!\times\! J_{\perp} $,\\
- Dirac's inspired point form: 
$t^2-\vec{x}^2 =\tau,\;\;\;\xi^0=u^0=1, \;\;\;  \;
\vec{\xi}=\vec{u}\,,$\cite{Desplanques:2004rd}\\
\hspace*{6mm}where $\vec{u}$ is a unit vector that points to any direction,
consistently with the \hspace*{6mm}absence of a particular 3-direction 
on a hyperboloid ($\xi^2=0$);\\ 
\hspace*{1cm}dynamical: $P^0,\;\vec{P}, \;\;\;$ kinematical:
$\vec{K},\;\vec{J}\,$.\\
An ``instant-form'' approach ``which displays the symmetry 
properties inherently present in the point-form'' one has been 
proposed.\cite{Bakamjian:1961} The Poincar\'e group generators, $P^{\mu}$ 
and  $M^{\mu \nu}$, have respectively a dynamical and a kinematical 
character, as for the Dirac's point form. 
However, as noticed by Sokolov,\cite{Sokolov:1985jv} it implies physics 
described on an hyperplane perpendicular to the velocity of the system under 
consideration (hypersurface $v \cdot x=\tau$). 
It therefore differs from the Dirac's one. This 
``point form'', which has been referred to in many recent 
applications,\cite{Wagenbrunn:2000es,Allen:1998hb,Allen:2000ge,Desplanques:2001zw,Amghar:2003tx} 
evidences specific features. Contrary to the other approaches mentioned above, 
the 4-vector, $\xi^{\mu}$, depends on the properties of the system 
($\xi^{\mu} \propto P^{\mu}$). 
This approach is also on a different 
footing with other respects.\cite{Desplanques:2004rd} 

For the mass operator, we refer to an equation used in our previous 
works\cite{Amghar:2002jx,Desplanques:2004rd,Desplanques:2004sp} with
appropriate changes due to the 1/2-spin of the 
constituents.\cite{Amghar:2003tx} For our purpose,
which is mainly to compare different approaches between themselves rather 
than to experiment, we include in the interaction a confining potential with
string tension, $\sigma_{s.t.}= 1\,$GeV/fm and 
a gluon exchange one with strength $\alpha_s=0.35$. This last contribution 
is of relevance to test the ability of RQM approaches in reproducing 
the expected QCD asymptotic behavior of form factors. As been 
noticed,\cite{Alabiso:1974sg} this behavior is closely related to the 
most singular part of the interaction at short distances. 
The quark and pion masses are taken as $m_q=0.3\,$GeV and $M_{\pi}=0.14\,$GeV.

Expressions of the single-particle contribution to form factors in the spinless 
case have been given elsewhere
(see for instance Ref. 6). They can be expressed
as an integral over the spectator-particle momentum. Interestingly, 
they  take  a unique form in most cases, which allows one to
discard major biases in the comparison of different approaches. Their derivation 
supposes to express the momenta of the constituents, $\vec{p}_{i,f}$ and 
$\vec{p}$ in Fig. \ref{fig1} in terms of the total momentum, $\vec{P}$,
the internal variable appearing in the mass operator, $\vec{k}$, and the
4-vector, $\xi^{\mu}$. The relation of $\vec{p}\,$'s to the $\vec{k}$ variable
assumes a Lorentz-type  transformation while fulfilling Eq. (\ref{eq1}). It is nothing but
the Bakamjian-Thomas one in a particular case.\cite{Bakamjian:1953kh} 
\begin{figure}[htb]
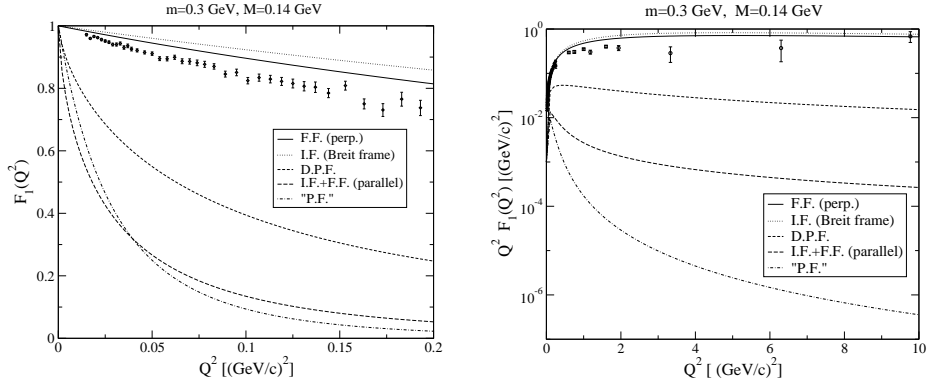
 
\epsfig{file=figpis.eps, width = 5.8cm} 
\hspace*{4mm} \epsfig{file=figpiS.eps, width = 5.8cm}
\caption{Pion charge form factor at low and high $Q^2$ together with 
experimental data 
\label{FF1}}
\end{figure} 
%%%%%%%%%%%%%%%%%%%%%%%-33333333333333-%%%%%%%%%%%%%%%%%%%%%%%%%%%%%%%%%%%%%%%
%%%%%%%%%%%%%%%%%%%%%%%%%%%%%%%%%%%%%%%%%%%%%%%%%%%%%%%%%%%%%%%%%%%%%%%%%%%%%%%
%%%%%%%%%%%%%%%%%%%%%%%%%%%%%%%%%%%%%%%%%%%%%%%%%%%%%%%%%%%%%%%%%%%%%%%%%%%%%%%
%%%%%%%%%%%%%%%%%%%%%%%%%%%%%%%%%%%%%%%%%%%%%%%%%%%%%%%%%%%%%%%%%%%%%%%%%%%%%%
\section{Results for the charge and scalar pion form factors}
The pion has two form factors: the charge one, $F_1(Q^2)$, for which 
measurements are available\cite{Bebek:1978pe,Amendolia:1986wj,Volmer:2000ek} 
and a scalar one, $F_0(Q^2)$, which, in absence of an appropriate probe, 
is unknown but can be nevertheless useful for a comparison of different 
approaches. The low and high $Q^2$ behaviors of $F_1(Q^2)$, in relation 
with the charge radius or the asymptotic behavior, are of special interest. 
Moreover, as the instant- and front-form form factors are not Lorentz 
invariant, they can be considered for various kinematical configurations. 
Besides the standard ones (respectively Breit frame and $q^+=0$), 
we consider both of them for a parallel kinematics 
and $|\vec{P}_i+\vec{P}_f| \rightarrow \infty$, where they coincide. We also
consider results in two point-form approaches, which contrary to the other
forms, are Lorentz invariant. 

\begin{figure}[htb]
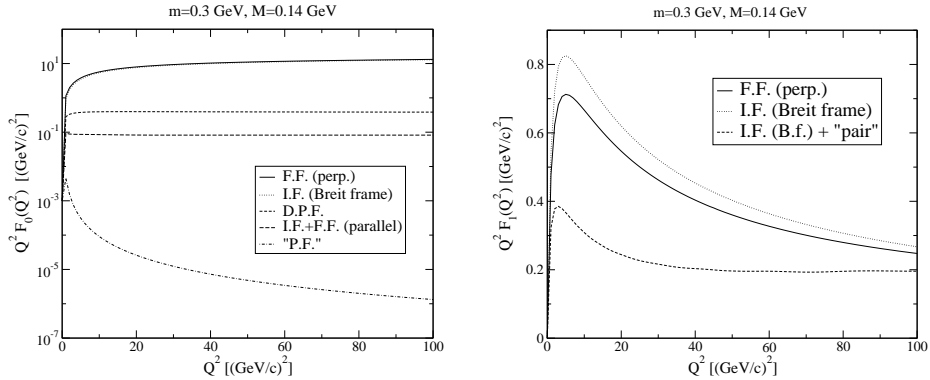
 
\epsfig{file=figpi0SS.eps, width = 5.8cm} 
\hspace*{4mm} \epsfig{file=pairSS.eps, width = 5.8cm}
\caption{Pion scalar and charge form factor in the asymptotic regime\label{FF2}}
\end{figure}

Results for $F_1(Q^2)$ are presented in Fig. \ref{FF1}. They clearly fall 
into two sets: the standard instant- and front-form form factors that 
are relatively close to experiment and the other ones that are far apart. 
Looking in detail at these last ones, it is found that they roughly depend 
on the momentum transfer $Q$ through the quantity $(2\bar{e_k}/M_{\pi})\,Q$, 
hence a charge radius scaling like the inverse of the pion mass, 
which explains the rapid fall off of the corresponding form
factors at low $Q^2$ (a  rapid fall off is also found in truncated field-theory
calculations\cite{Bakker:2000pk,Simula:2002vm,deMelo:2002yq}). 
At higher $Q^2$, it sounds that the $Q^{-2}$ asymptotic 
behavior is reached. Actually, examination of the charge form factor 
at much higher $Q^2$ indicates that the behavior is $Q^{-4}$ 
(see Fig. \ref{FF2}). The consideration of the scalar form factor, $F_0(Q^2)$, 
is especially useful here. As Fig. \ref{FF2} shows, this form factor 
has the QCD-expected $Q^{-2}$ asymptotic behavior, indicating that there 
is nothing wrong with the solution of the mass operator we used. In order 
to get the right power-law behavior for the charge form factor, we considered 
the contribution of two-body currents (pair-type term). This one, determined 
so that to reproduce the full Born amplitude,\cite{Desplanques:2003nk} 
has been calculated in the 
instant form. As Fig. \ref{FF2} shows, it provides the right $Q^{-2}$ 
asymptotic behavior. Moreover, the coefficient 
has the expected expression (up to a numerical factor).
%

%%%%%%%%%%%%%%%%%%%%%%%%%-444444444444444-%%%%%%%%%%%%%%%%%%%%%%%%%%%%%%
%%%%%%%%%%%%%%%%%%%%%%%%%%%%%%%%%%%%%%%%%%%%%%%%%%%%%%%%%%%%%%%%%%%%%%%%%%%%%%%
%%%%%%%%%%%%%%%%%%%%%%%%%%%%%%%%%%%%%%%%%%%%%%%%%%%%%%%%%%%%%%%%%%%%%%%%%%%%%%
%%%%%%%%%%%%%%%%%%%%%%%%%%%%%%%%%%%%%%%%%%%%%%%%%%%%%%%%%%%%%%%%%%%%%%%%%%%%%
\section{Conclusion and prospect}
%\begin{figure}[htb] 
% \epsfig{file=kine2.eps, width= 8cm} 
%\end{figure}

The examination of the pion charge form factor calculated from a 
single-particle current in different RQM approaches shows unambiguously 
that results fall into ``good'' and ``bad'' ones. The conclusion is not 
to be affected by refining the physical description as the discrepancy 
in the last case reaches huge factors. It fully confirms the conclusions 
achieved in the spinless case whose physical description 
is simpler.\cite{Desplanques:2004sp} The QCD asymptotic behavior is obtained 
from two-body currents. 

Lorentz invariance is often advertised as a  validity criterion  of some
approach. This view is not however supported by present point-form results, 
which explicitly evidence the above invariance property. Moreover, 
the violation of Lorentz invariance, as measured from the rather small 
discrepancy between the standard instant- and front-form results, 
does not seem to be necessarily large. Another criterion has 
therefore to be found. In a field-theory approach, the 4-momentum 
is conserved at the vertex representing the interaction of constituents 
with the external probe. This cannot be generally fulfilled 
in RQM approaches at the operator level (unless many-body currents 
are considered). One can however require that the property be verified 
at the level of the matrix element. Considering this weaker argument, 
it allows one to account for the observed discrimination of results 
into ``good'' and ``bad'' ones. As the 4-momentum conservation stems 
from Poincar\'e space-time translation invariance, fulfilling 
this property could be the relevant criterion. The above invariance 
may also be the important symmetry whose violation is suggested 
by the peculiar behavior of some form factors in the limit 
of a zero-mass system.\cite{Desplanques:2004sp,Desplanques:2003nk}

Poincar\'e space-time translation invariance implies relations 
such as:\cite{Lev:1993}
\begin{equation}
\Big[ P^{\mu}\;,\; J^{\nu}(x)\Big]=-i\;\partial^{\mu}\,J^{\nu}(x), 
\end{equation}
which can be used for a quantitative check. Considering a single-particle 
current, it is found that the equality is satisfied at the matrix-element 
level for the standard instant- and front-form results. It is violated 
in all the other cases. Skipping details, one finds that the l.h.s. 
and  r.h.s respectively involve the quantities  $Q$ 
and $(2\bar{e_k}/M_{\pi})\;Q$. The extra factor at the r.h.s., 
$(2\bar{e_k}/M_{\pi})$, which is the same as the one 
explaining the discrepancy between the ``good'' and ``bad'' 
form factors in Fig. \ref{FF1}, provides a measure of the violation 
of Poincar\'e space-time translation invariance. To get rid of it, 
interaction currents should be considered. Schematically, their effect 
could combine with the kinetic energy term, $ 2\,\bar{e_k}$, 
at the numerator of the above factor so that the overall factor 
be 1 (using $2\,\bar{e_k}+V=M_{\pi}$).\cite{Desplanques:2003nk}

Coming back to the motivation of the present work, it is noticed that 
the success of the point-form description of the nucleon form factors 
is mostly due to a factor similar to the above one. As a more complete
calculation is expected to remove this factor, a situation similar 
to the standard front-form calculation should be recovered. 
We believe it is doubly fortunate. 
The description of the nucleon form factor could now incorporate 
the well known vector-meson dominance phenomeno\-lo\-gy. The difficulty
to reconcile the point-form descriptions of the nucleon 
and pion form factors, respectively good\cite{Wagenbrunn:2000es} 
and bad,\cite{Amghar:2003tx} vanishes.

While Lorentz invariance has often been advocated in calculating 
form factors, Poincar\'e space-time translation invariance could 
be a more relevant property.
%%%%%%%%%%%%%%%%%%%%%%%%%%%%%%%%%%%%%%%%%%%%%%%%%%%%%%%%%%%%
% Doing Acknowledgement						     %
%%%%%%%%%%%%%%%%%%%%%%%%%%%%%%%%%%%%%%%%%%%%%%%%%%%%%%%%%%%%

\section*{Acknowledgements}

We are very grateful to A. Amghar, T. Melde, S. Noguera and L. Theu{\ss}l 
for discussions or comments at various steps of the present work.

%%%%%%%%%%%%%%%%%%%%%%%%%%%%%%%%%%%%%%%%%%%%%%%%%%%%%%%%%%%%
% Doing references:	                   	           %
%%%%%%%%%%%%%%%%%%%%%%%%%%%%%%%%%%%%%%%%%%%%%%%%%%%%%%%%%%%%

\end{document}